\documentclass[11pt,showpacs,preprintnumbers,amsmath,amssymb]{revtex4}
\usepackage{graphicx}% Include figure files
\begin{document}

\title{ Low-Temperature Quasi-Equilibrium States in the Hydrogen Atom}

\vskip \baselineskip

\author{N.~M. Oliveira-Neto$^{1}$}
\thanks{E-mail address: nmon@cbpf.br}

\author{E.~M.~F. Curado$^{1}$}
\thanks{E-mail address: evaldo@cbpf.br}

\author{F.~D. Nobre$^{1,2}$}
\thanks{Corresponding author: E-mail address: nobre@dfte.ufrn.br}

\author{M.~A. Rego-Monteiro$^{1}$}
\thanks{E-mail address: regomont@cbpf.br}

\address{
$^{1}$Centro Brasileiro de Pesquisas F\'{\i}sicas \\
Rua Xavier Sigaud 150 \\
22290-180 \hspace{5mm} Rio de Janeiro - RJ \hspace{5mm} Brazil \\
$^{2}$Departamento de F\'{\i}sica Te\'orica e Experimental \\
Universidade Federal do Rio Grande do Norte \\
Campus Universit\'{a}rio -- Caixa Postal 1641 \\
59072-970 \hspace{5mm} Natal - Rio Grande do Norte \hspace{5mm} Brazil}

\date{\today}

%\newpage
\begin{abstract}
\noindent
The dynamics of the approach to equilibrium of the hydrogen atom is 
investigated numerically through
a Monte Carlo procedure. We show that, before approaching ionization, 
the hydrogen atom may live in a quasi-equilibrium state, characterized 
by aging, 
whose duration increases exponentially as the temperatures decreases. 
By analyzing the quasi-equilibrium 
state, we compute averages of physical quantities for the hydrogen atom. 
We have introduced an analytic approach that fits satisfactorily the
numerical estimates for low temperatures.
Although the present analysis 
is expected to hold for energies typically up to 
$6 \% $ of the ionization energy, it works well for temperatures 
as high as $10^{4}$ K.

%\vskip \baselineskip
\vspace{1cm}
%\medskip


\end{abstract}
\maketitle

\newpage

\noindent
{\large\bf 1. \quad Introduction}

\vskip \baselineskip

Obtaining equilibrium properties for the hydrogen atom in free space,
through standard Boltzmann-Gibbs (BG) statistical mechanics, is
troublesome, since the partition function diverges for any finite
temperature. This occurs mostly in systems of composite particles, 
which are characterized by upper bounds, preceded by a quasi-continuum 
of energy levels, in their energy spectra. Each of the levels in the
quasi-continuum yields a small contribution for the partition function;
the divergence arises since there are, in principle, an infinite number 
of such levels. Essentially, this reflects the 
fact that composite particles always ionize at any finite temperature. 
Due to these difficulties, such systems are never discussed in standard
statistical-mechanics textbooks (for an exception of this, see
Ref. \cite{fowler}). 

However, the presence of long-living nonionized hydrogen atoms in
galaxy peripheries and in intergalactic media, at low temperatures, is
undeniable. Certainly, their density must be very low in such a
media, in order to avoid combination 
(${\rm H} + {\rm H} \rightarrow {\rm H}_{2}$), and they should spend a long
time in their nonionized state, before reaching ionization. 
These long-living quasi-equlibrium states represent the
main point we explore, on theoretical grounds, in the present work. 

The specific heat of the hydrogen atom has been calculated recently 
\cite{liacir} through the generalized statistical-mechanics formalism
\cite{tsallisbjp,abeokamoto,grigolini} 
that emerged from Tsallis's generalization of
the BG entropy \cite{tsallis88}. Within this approach, the specific heat
was computed for certain values of the
entropic index $q<1$ (notice that $q=1$ corresponds to the standard 
BG formalism). Even within such a formalism, the hydrogen-atom specific heat
presents several anomalies, like divergences, cusps, and discontinuities in
its derivative \cite{liacir}.

Herein, we choose a different approach for analyzing this problem, as
described next.

(i) We investigate the
dynamical behavior of the hydrogen atom by applying a standard 
Monte Carlo method,
in which the probability for
jumping between states is based on the BG weight. As expected, for any
finite temperature, the simulation always carries the system towards
ionization, after some time. However, it is shown that 
the system may live 
in a quasi-equilibrium state, characterized by a slowly varying value of the
average energy, before 
approaching its maximum-energy state. The time that the system remains on
such a quasi-equilibrium state increases for lowering temperatures. 

(ii) We propose a modified dynamics that prevents the system from
reaching ionization and whose results coincide, during some time
(essentially when the system lies in its low-temperature quasi-equilibrium state),
with those of the standard dynamics. 
Therefore, for low temperatures, depending on time scale of interest, 
the modified dynamics may reflect the
correct dynamical behavior of the system. 
The advantage of the
modified dynamics is that its corresponding partition function, 
associated to the 
statistical weight that generated it, is finite and may be calculated
exactly. 

(iii) Since the quasi-equilibrium state
may present a long duration for low temperatures, herein we will
consider it as
an effective equilibrium. Therefore, if the results obtained from the 
standard and modified dynamics coincide within such a quasi-equilibrium state, 
one may use the later formalism in
such a way to compute thermodynamic properties.

(iv) We introduce a modified regularized partition function, by extracting
the divergence of the BG partition function, for a hydrogen atom, and
compare the corresponding internal energy and specific heat with the
results obtained from the numerical simulations. 

(v) Obviously, the procedures described above may work as good
approximations for low energies, but should fail for increasing energies. 
Since our energies are always measured with respect to the
corresponding ionization energy, we show that our approximations work well
for energies that correspond to temperatures much higher 
than room temperature. 

In the next section we investigate the dynamical behavior of the hydrogen
atom within a standard Monte Carlo framework. In section 3 we  introduce
a modified dynamics and apply it for the hydrogen atom. In section 4 we
calculate a regularized partition function, related to the modified
dynamics introduced in section 3, and compute some associated
thermodynamic quantities for the hydrogen atom. In section 5 we propose two
possible physical realizations for the quasi-equilibrium states 
presented herein. 
Finally, in section 6 we present our conclusions. 

\vskip 2 \baselineskip
\noindent
{\large\bf 2. \quad The Hydrogen Atom within a Standard Monte 
Carlo Procedure}

\vskip \baselineskip

Let us consider a hydrogen atom with its well-known energy spectrum
(see, e.g., ref. \cite{cohen}, page 790),  

\vspace{-5mm}

$$
E_{n} = R \left( 1 - {1 \over n^{2}} \right) \qquad (n=1,2, \cdots ),   
\eqno(1) 
$$

%\vskip \baselineskip
\noindent
where $R$ is the Rydberg constant [$R=13.6058$ eV, or 
$R=2.18 \times 10^{-18}$ J, which corresponds to a temperature
$(R/k_{B})=1.579 \times 10^{5}$ K], 
and we have chosen the ground state to have
zero energy. As $n$ increases, one has a quasi-continuum of energy levels,
and in the limit $n \rightarrow \infty$ the atom ionizes
(with an ionization energy $E^{*}=R$), in such a way that the gap 
separating the ground-state and ionization energies is $R$. 
A transition between the ground state and the first-excited state 
costs most of
the energy of the gap, i.e., $3R/4$, whereas all further jumps occur in
the range $R/4$. Therefore, for low temperatures, the hydrogen atom is
expected to remain a long time in the lowest-energy states (mostly in the
ground state). However, as the quantum number $n$ increases, the energy
cost for jumps between nearest-neighbor levels decreases. Therefore,
once the system has reached a state characterized by a
large quantum number $n$, transitions to higher-energy levels cost 
very little 
energy, in such a way that after a long time, the hydrogen atom will
ionize. The ionization will always occur for any finite temperature; one of
the questions we address in the present work is {\it how long} does the 
system remains in the quasi-equilibrium state (characterized by an average
energy close to the ground-state energy). 

It is important to remind that the above energy spectrum is degenerate
(see, e.g., ref. \cite{cohen}, page 798), in such a way that a
given energy level $E_{n}$ presents a total 
degeneracy $g_{n}=n^{2}$. Therefore, a precise analysis of this problem
should take into account the degeneracy of the energy levels, as well 
as of the spins. However, in the present 
investigation, only the lowest energy levels will
contribute most significantly, in such a way that the effects of the
degeneracy will not change qualitatively our results. Consequently, 
we will consider herein, $g_{n}=1 \ (\forall n)$, for simplicity.  

We have investigated the dynamical behavior of the hydrogen atom
through a Monte Carlo procedure \cite{garrod}. The 
probabilities $P(n+1 \rightarrow n)$ and $P(n \rightarrow n+1)$,
for transitions between the states characterized by quantum numbers $n$ and 
$n+1$, satisfying the detailed-balance condition and constructed by 
using the standard BG weight, are given by

\vspace{-5mm}

$$
P(n+1 \rightarrow n) = c~, \qquad 
P(n \rightarrow n+1) = c \ \exp [-\beta(E_{n+1}-E_{n})]~,   
\eqno(2) 
$$

%\vskip \baselineskip
\noindent
where $\beta = (k_{B}T)^{-1}$ and $c$ is an arbitrary constant 
($0 < c \leq 1/2$). 
In order to
implement the dynamical evolution, a uniform random number $z$ 
($z \in [0,1]$) must be generated at each Monte Carlo step (which will be
adopted as our unit of time). For a system in a state characterized by the
quantum number $n$ at time $t$, transitions between
states are performed (or not), depending on the value of $z(t)$, 
according to the following rules:

(i) If $z(t) \leq P(n \rightarrow n+1)$, perform the jump $n \rightarrow n+1$;

(ii) If $P(n \rightarrow n+1) < z(t) 
\leq [P(n \rightarrow n+1)+ P(n \rightarrow n-1)]$, 
perform the jump $n \rightarrow n-1$;

(iii) Else, remain on level $n$.

\noindent
One may easily see that the constant $c$ is proportional to the 
probability for no jumps [rule (iii)]. In the following results we have
used $c=1/2$, although we have verified that other choices for this
constant ($0<c<1/2$) did not change qualitatively our results. Below, 
$\langle \ \rangle $ correspond 
to averages over $N_{s}$ distinct samples,
i.e., different sequences of random numbers. We have considered two distinct
initial conditions in our simulations, which will be reffered to, herein,
as conditions 1 and 2, corresponding 
to all samples starting with the quantum number $n=1$ (condition 1), and  
all samples starting with $n=2$ (condition 2). Although, in some cases, the
results obtained by using these two initial conditions seem to be
different, they approach each other in the limit 
$N_{s} \rightarrow \infty$ (except, of course, in the transient regimes,
before reaching the quasi-equilibrium state). 

%%%%%%%%%%%%%%%%%%%%%%%%%%%%%%%%%%%%%%%%%%%%%%%%%%%%%%%%%%%%%%%%%%%%%%%%%%%
\begin{figure}
\begin{center}
\includegraphics[angle=0,scale=0.4]{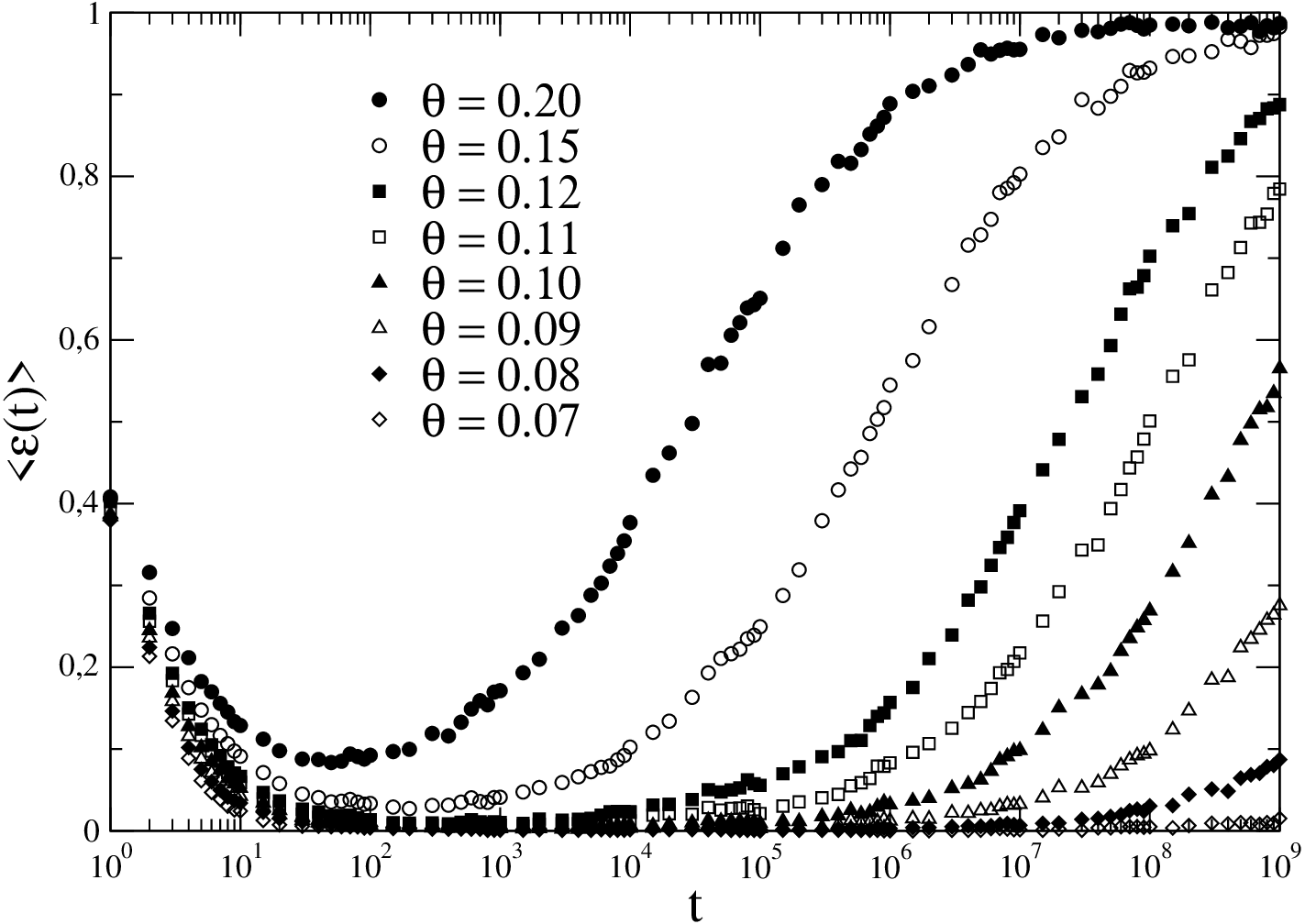}
\end{center}
\caption{\small
The time evolution of the average dimensionless energy (energy in units of 
the Rydberg constant), for the hydrogen atom, 
is represented for several relative temperatures 
($\theta = k_{B}T/R$). Time is measured in Monte Carlo steps.}
\label{fig1}
\end{figure}
%%%%%%%%%%%%%%%%%%%%%%%%%%%%%%%%%%%%%%%%%%%%%%%%%%%%%%%%%%%%%%%%%%%%%%%%%%%

In Fig. 1 we present the time evolution of the average dimensionless
energy, $\langle \varepsilon (t) \rangle$ 
[$\varepsilon_{n}=E_{n}/R$], 
for several values of the relative-temperature variable,
$\theta =(k_{B}T)/R$, i.e., the ratio of the temperature
with respect to the Rydberg constant.
Our simulations were carried up to a maximum time $t_{\rm max}=10^{9}$,
whereas for the averages we have considered $N_{s}=2000$ samples with the
initial condition 2. Our plots
exhibit a general tendency for increasing the average energy, towards the
ionization energy, after some time. This reflects the fact that the
hydrogen atom always ionize for any finite temperature. 
However, the interesting
effect noticed herein is the presence of a quasi-equilibrium state,
characterized by a slowly varying average energy, before the approach 
to ionization. Such a quasi-equilibrium state 
may present a long duration for low values of $\theta$, as shown in 
Fig. 1. 

%%%%%%%%%%%%%%%%%%%%%%%%%%%%%%%%%%%%%%%%%%%%%%%%%%%%%%%%%%%%%%%%%%%%%%%%%%%
\begin{figure}
\begin{center}
\includegraphics[angle=0,scale=0.4]{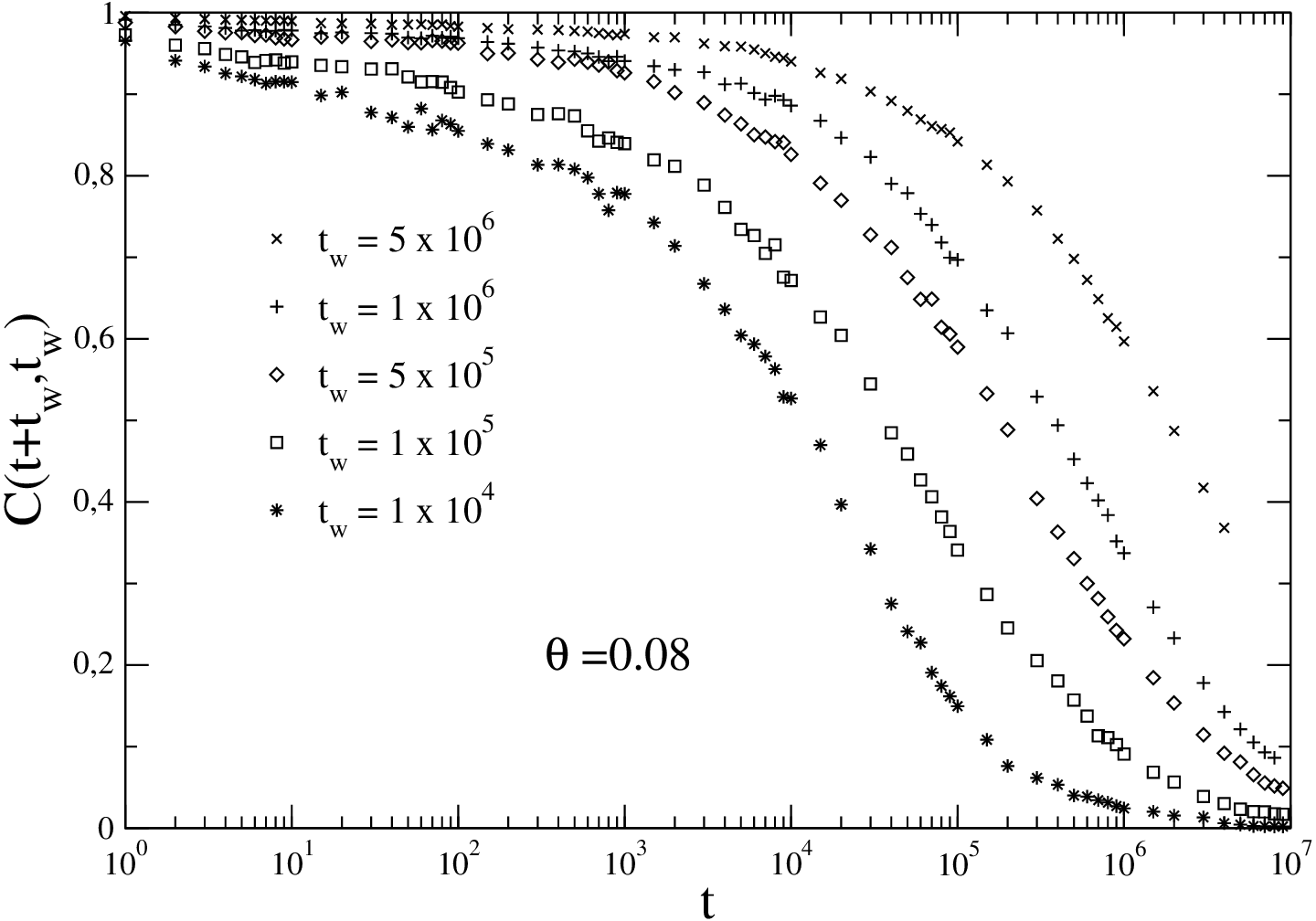}
\end{center}
\caption{\small
The two-time autocorrelation function $C(t+t_{w},t_{w})$,
for the hydrogen atom, is represented 
against time, for a given value of the relative temperature and 
several waiting times $t_{w}$. Time is measured in Monte Carlo steps.}
\label{fig2}
\end{figure}
%%%%%%%%%%%%%%%%%%%%%%%%%%%%%%%%%%%%%%%%%%%%%%%%%%%%%%%%%%%%%%%%%%%%%%%%%%%

It is important to notice that such a state presents quite
nontrivial behavior, e.g. it is characterized by aging, as observed on a
similar system \cite{neto04}. In order to see this effect, let us define the
two-time autocorrelation function \cite{cugliandolo},
 
\vspace{-5mm}

$$
C(t+t_{w},t_{w}) = { \langle \varepsilon (t+t_{w})  
\varepsilon (t_{w}) \rangle
- \langle \varepsilon (t+t_{w}) \rangle \langle  \varepsilon (t_{w}) \rangle
\over \sigma (t+t_{w})  \sigma (t_{w}) },
\eqno(3 {\rm a})
$$

%\vskip \baselineskip
\noindent
where $t_{w}$ represents the well-known ``waiting time'', and  

\vspace{-5mm}

$$
\sigma (t) = \left[ \langle (\varepsilon (t))^{2} \rangle 
- \langle \varepsilon (t) \rangle ^{2} \right]^{1/2}. 
\eqno(3 {\rm b})
$$

\noindent
In Fig. 2 we exhibit such a correlation function 
for a value of the relative temperature $\theta = 0.08$ 
and several waiting times. The initial conditions are the same as in
Fig. 1, but now, the averages that appear in Eqs. (3) 
were taken over $N_{s}=350000$ samples. 
The waiting times considered were
chosen in such a way to ensure that the correlation function 
$C(t+t_{w},t_{w})$ was evaluated with the system on the quasi-equilibrium
state (which is typically inside the time range 
$10^{2} \rightarrow 10^{8}$,
as shown in Fig. 1). 
Clearly, there is a dependence on the waiting time typical of the aging 
effect \cite{cugliandolo}. 

\vskip 2 \baselineskip
\noindent
{\large\bf 3. \quad The Hydrogen Atom and the Modified Monte 
Carlo Procedure}

\vskip \baselineskip

We have also considered the evolution of the system under a 
modified dynamics, which satisfies detailed balance, but 
prevents the approach to ionization
(the justification for that will become clear later on). The jumping
probabilities are given by

\vspace{-5mm}

$$
P(n+1 \rightarrow n) = c~, \qquad 
P(n \rightarrow n+1) = c \ { \exp (-\beta E_{n+1}) - \exp (-\beta R)
\over \exp (-\beta E_{n}) - \exp (-\beta R) }~.   
\eqno(4) 
$$

%\vskip \baselineskip
\noindent
In Fig. 3 we compare the results for the 
time evolution of the average dimensionless
energy obtained by using the
standard dynamics [Eq. (2)] with those obtained through the 
modified dynamics [Eq. (4)], for typical values of the relative
temperature. 
The conditions for the simulations in Fig. 3 are the same as those
of Fig. 1, which correspond to $t_{\rm max}=10^{9}$ and $N_{s}=2000$
(initial condition 2). 
One observes that the
agreement between the two dynamical procedures persists during a given
time (essentially within the quasi-equilibrium state), 
that increases for lowering values of the temperature. 
Within the
modified dynamics, which was constructed in such a way  
to avoid the system from reaching 
ionization, the system remains on a 
quasi-equilibrium state forever. 

%%%%%%%%%%%%%%%%%%%%%%%%%%%%%%%%%%%%%%%%%%%%%%%%%%%%%%%%%%%%%%%%%%%%%%%%%%%
\begin{figure}
\begin{center}
\includegraphics[angle=0,scale=0.4]{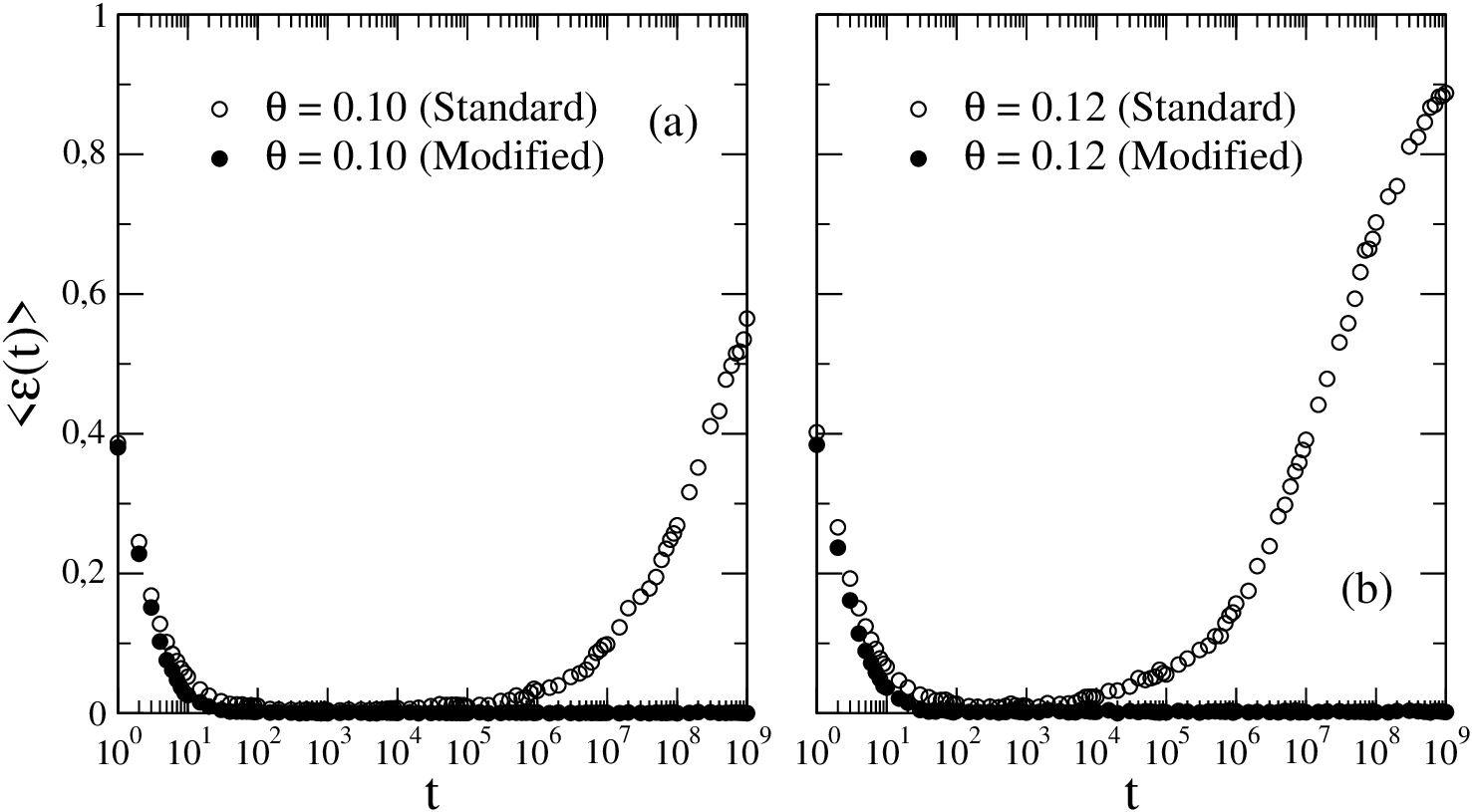}
\end{center}
\caption{\small
The time evolution (in Monte Carlo steps) of the average dimensionless 
energy (energy in units of 
the Rydberg constant), for the hydrogen atom, 
is represented for two relative temperatures 
($\theta = k_{B}T/R$), in the case of the standard (empty symbols) and
modified dynamics (full symbols).}
\label{fig3}
\end{figure}
%%%%%%%%%%%%%%%%%%%%%%%%%%%%%%%%%%%%%%%%%%%%%%%%%%%%%%%%%%%%%%%%%%%%%%%%%%%

Let us define the duration of the quasi-equilibrium state, 
$t_{\rm MS}(\theta)$, 
as the time during which the system remains on such a state, within
the standard dynamics, by keeping the absolute value of the difference 
between
the average dimensionless energies computed from the standard and modified
dynamics less than a given value $\delta$. Although the choice of $\delta$
may be arbitrary, one expects the corresponding law followed by
$t_{\rm MS}(\theta)$ to be independent of this particular choice. 
In the present analysis we estimated $t_{\rm MS}(\theta)$ by considering 
several values of
$\theta$ from $0.08$ up to $0.12$, by imposing that $\delta$ does not
exceed $0.05$. Our data fit well the exponential law,

\vspace{-5mm}

$$
t_{\rm MS}(\theta) \sim \exp (b/\theta) = \exp [b R/(k_{B}T)] 
\qquad (b = 1.95 \pm 0.06),   
\eqno(5) 
$$

%\vskip \baselineskip
\noindent
which implies $t_{\rm MS}(\theta) \rightarrow \infty$ when 
$T \rightarrow 0$.
Essentially, the duration of the quasi-equilibrium state follows an Arrenhius
law, typical of the Kramers' escape problem in chemical reactions
\cite{kramers,hanggi}, where the system may remain for a long time 
in a quasi-stationary state, before overcoming the potential barrier
associated with the reaction. To our knowledge, this is the first time that
such a behavior has been associated with the dynamics of the hydrogen atom. 

Therefore, for very low temperatures and
depending on time scale of interest, the modified dynamics may reflect the
correct dynamical behavior of the system. The advantage of the modified
approach is that the corresponding partition function, associated to the 
statistical weight that generated Eq. (4), is finite -- contrary to the
one associated with the standard dynamics -- and may be calculated
exactly. This will be done in the next section. 

\vskip 2 \baselineskip
\noindent
{\large\bf 4. \quad The Regularized Partition Function and 
Associated Thermodynamic Functions}

\vskip \baselineskip

In the present section we shall calculate the partition function defined
through the statistical weight related to the modified dynamics, introduced
above. 
Let us address this point by considering 
in detail the divergence of the partition function
associated with the energy spectrum of Eq. (1),

\vspace{-5mm}

$$
Z = \sum _{n=1}^{\infty} {\rm exp}(-\beta E_{n}) = {\rm exp}(- \beta R)
\sum _{n=1}^{\infty} {\rm exp}[\beta R /(n^{2})]~.   
\eqno(6) 
$$

%\vskip \baselineskip
\noindent
The equation above may still be written as

\vspace{-5mm}

$$
Z = {\rm exp}(- \beta R) \ \lim _{n^{*} \rightarrow \infty} \
\sum _{n=1}^{n^{*}} {\rm exp}[\beta R /(n^{2})]~,   
\eqno(7) 
$$

%\vskip \baselineskip
\noindent
where $n^{*}$ is an appropriated cutoff in the quantum number, 
that will be taken to infinite later on. One has that

\vspace{5mm}

\setcounter{enumi}{8}
\renewcommand{\theequation}{\arabic{enumi}}
\begin{eqnarray}
{\rm exp}(\beta R) Z & = & \lim _{n^{*} \rightarrow \infty} \
\sum _{n=1}^{n^{*}} \sum _{m=0}^{\infty}{(\beta R)^{m} \over m!}
{1 \over n^{2m}}
= \lim _{n^{*} \rightarrow \infty} \
\sum _{n=1}^{n^{*}} \left[ 1 
+ \sum _{m=1}^{\infty}{(\beta R)^{m} \over m!}
{1 \over n^{2m}} \right] \nonumber \\
\nonumber \\
& = & \lim _{n^{*} \rightarrow \infty} \
\left[ n^{*} + \sum _{m=1}^{\infty} H_{n^{*},2m} \ 
{(\beta R)^{m} \over m!} \right]~,
\end{eqnarray}

\vspace{5mm}

%\vskip \baselineskip
\noindent
where $H_{n^{*},2m}=\sum _{n=1}^{n^{*}} 1/(n^{2m})$ are the harmonic
numbers of order $2m$ \cite{graham}. The limits 
$n^{*} \rightarrow \infty$ of the harmonic numbers are well-defined,
leading to finite coefficients,
$B_{2m} = \lim _{n^{*} \rightarrow \infty} H_{n^{*},2m}$. One has that
$B_{2} = \pi ^{2}/6 = 1.64493...$, 
$B_{4} = \pi ^{4}/90 = 1.08232...$, 
$B_{6} = \pi ^{6}/945 = 1.01734...$, in such a way that $B_{2m}$ converges to
unit for increasing values of $m$, e.g., $B_{16}=1.00002...$. Therefore, one
gets

\vspace{-5mm}

$$
Z = {\rm exp}(- \beta R) \left[ \lim _{n^{*} \rightarrow \infty}
(n^{*}) + \sum _{m=1}^{\infty} B_{2m} \ 
{(\beta R)^{m} \over m!} \right]~,   
\eqno(9) 
$$

%\vskip \baselineskip
\noindent
which shows a linear divergence with the quantum number. It should be 
stressed that the divergent contribution of the partition function comes
from a {\it single} term in the sum over $m$ of Eq. (8) (term $m=0$).
Let us now introduce a ``modified regularized partition function'', 

\vspace{-5mm}

$$
Z^{\prime} = Z - {\rm exp}(- \beta R) \lim _{n^{*} \rightarrow \infty}
(n^{*}) = {\rm exp}(- \beta R) \sum _{m=1}^{\infty} B_{2m} \ 
{(\beta R)^{m} \over m!}~,     
\eqno(10) 
$$

%\vskip \baselineskip
\noindent
which is finite. The regularized partition function defined above may be
written also as 

\vspace{-5mm}

$$
Z^{\prime} = \sum _{n=1}^{\infty} {\rm exp}(-\beta E_{n}) 
- \sum _{n=1}^{\infty} {\rm exp}(-\beta R) = 
\sum _{n=1}^{\infty} \left[ {\rm exp}(-\beta E_{n})
- {\rm exp}(-\beta R) \right]~,     
\eqno(11) 
$$

%\vskip \baselineskip
\noindent
where one identifies the statistical weight that leads to the 
jumping probabilities of Eq. (4).

Obviously, $Z$ and $Z^{\prime}$ are very distinct from one
another (actually, the difference between them diverges). However, 
for low temperatures, the hydrogen atom remains for 
a long time in its low-energy states. Therefore, for a given low 
temperature, if one considers the
corresponding quasi-equilibrium state as an effective equilibrium, 
for which the divergent term is not relevant, one may use 
$Z^{\prime}$ in order to calculate thermodynamic properties as
approximations. Within this formalism, the quantities that would correspond
to the internal energy and specific heat are given,
respectively, by

\vspace{-5mm}

\setcounter{enumi}{12}
\setcounter{equation}{0}
\renewcommand{\theequation}{\arabic{enumi}\alph{equation}}
\begin{eqnarray}
u^{\prime} = - R \ {\partial \ln Z^{\prime} \over \partial (\beta R)}
& = & -R \left\{ -1 + 
{\sum _{m=1}^{\infty}m (B_{2m}/m!) \ (\beta R)^{m-1} \over             
\sum _{m=1}^{\infty}(B_{2m}/m!) \ (\beta R)^{m} } \right\}~, \\          
\nonumber \\
{c^{\prime} \over k_{B}} = (\beta R)^{2} \ 
{\partial ^{2} \ln Z^{\prime} \over \partial (\beta R)^{2}}
& = & (\beta R)^{2} \ \left\{
{\sum _{m=1}^{\infty}m(m-1) (B_{2m}/m!) \ (\beta R)^{m-2} \over             
\sum _{m=1}^{\infty}(B_{2m}/m!) \ (\beta R)^{m} } \right. \nonumber \\
\nonumber \\   
& - & \left. 
{\left[ \sum _{m=1}^{\infty}m (B_{2m}/m!) \ (\beta R)^{m-1} \right]^{2}
\over             
\left[ \sum _{m=1}^{\infty}(B_{2m}/m!) \ (\beta R)^{m} \right]^{2}}
\right\}~,          
\end{eqnarray}

%\vskip \baselineskip
\noindent
where we keep the prime notations to remind that such thermodynamic
properties were calculated by using the regularized partition function
$Z^{\prime}$. 

%%%%%%%%%%%%%%%%%%%%%%%%%%%%%%%%%%%%%%%%%%%%%%%%%%%%%%%%%%%%%%%%%%%%%%%%%%%
\begin{figure}
\begin{center}
\includegraphics[angle=0,scale=0.4]{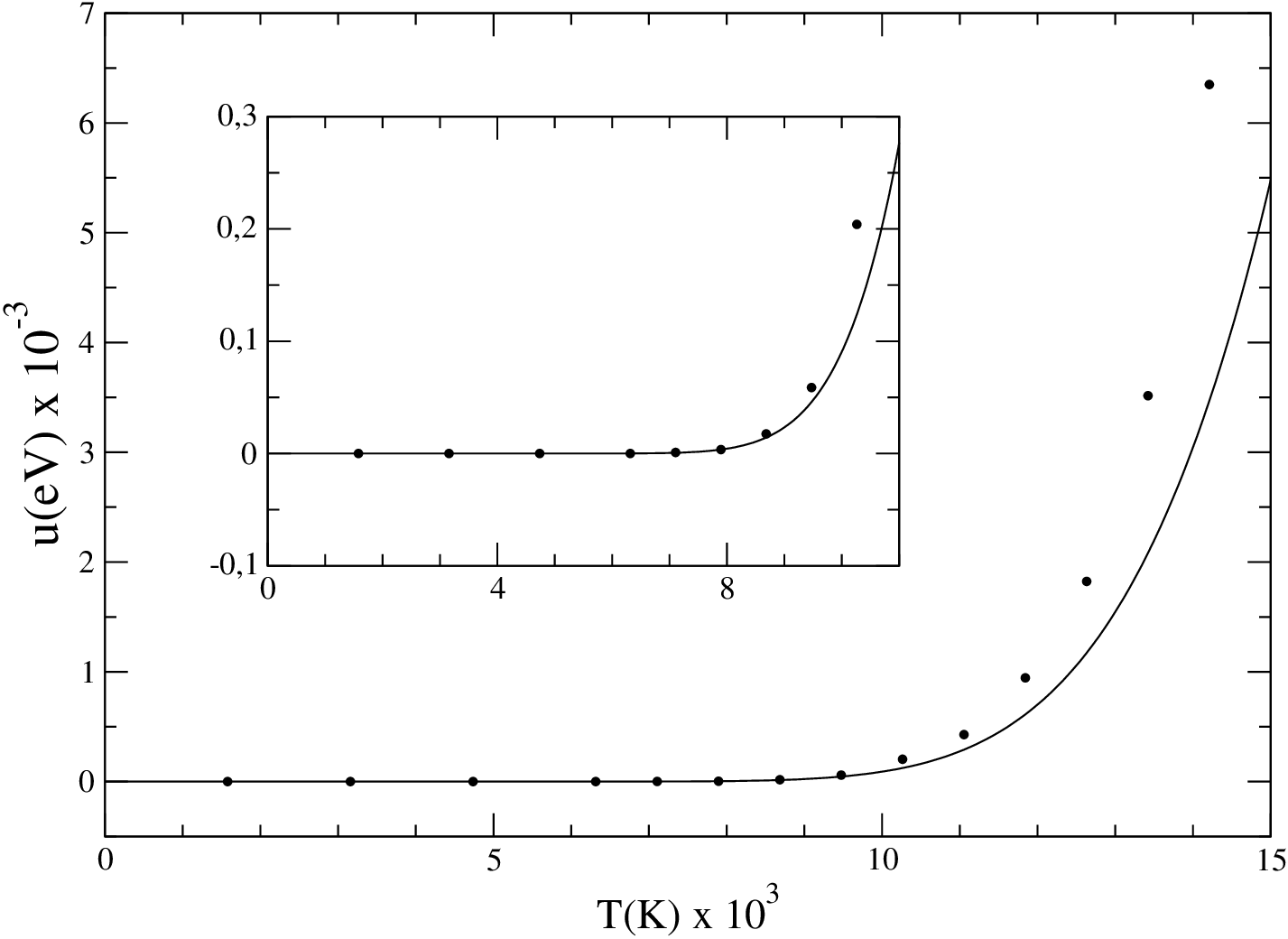}
\end{center}
\caption{\small
The internal energy of the hydrogen atom is calculated up to $T=15000$ K,
by two different approximations: (i) The analytic modified regularized 
partition
function formalism (full line); (ii) Standard Monte Carlo simulation (black
dots). The inset shows an amplification of these results in the
temperature range $T=0$ K $\rightarrow T=11000$ K.}  
\label{fig4}
\end{figure}
%%%%%%%%%%%%%%%%%%%%%%%%%%%%%%%%%%%%%%%%%%%%%%%%%%%%%%%%%%%%%%%%%%%%%%%%%%%

In Fig. 4 we exhibit the internal energy of the hydrogen
atom, as calculated from the above analytic expression [Eq. (12a)], 
and compare it with results of the 
numerical simulations. In the latter, we have computed time averages
over a time interval ranging from $t_{1}=200$ up to $t_{2}=400$ 
Monte Carlo steps of the standard-dynamics quasi-equilibrium
states, where the values at each time do already represent averages over 
$N_{s}=10^{6}$ samples, with the initial condition 1. 
Similar results may be obtained by starting the system with the initial
condition 2, although a larger computational effort (i.e., higher values of
$N_{s}$) may be necessary for a proper convergence towards the
low-temperature quasi-equilibrium state in this case. One observes a very good
agreement between the two approaches up to temperatures $10^{4}$ K. In
fact, the relative discrepancy between these procedures gets larger for
increasing temperatures, yielding the typical values of  0.004, for
$\theta=0.05$ ($T \approx 7895$ K) and 0.282, for
$\theta=0.06$ ($T \approx 9474$ K). 
A similar picture to the one shown in Fig. 4 holds for the specific heat of
the hydrogen atom, with relative discrepancies of the same order of
magnitude as those found for the internal energy. For the range of
temperatures over which one finds a {\it good agreement} 
between the two
approaches (typically from 0 K to $10^{4}$ K), 
{\it the long-living low-temperature quasi-equilibrium state} 
may be considered as an {\it effective equilibrium state}, 
for which the divergence of the partition
function in Eq. (9) has been removed, and one may compute 
effective thermodynamic
properties for the hydrogen atom from the regularized
partition function of Eq. (10). 
It is important to mention that the range of temperatures 
over which the present 
regularized formalism should be applied safely for the hydrogen atom 
goes {\it far beyond room temperature}.

\vskip 2 \baselineskip
\noindent
{\large\bf 5. \quad Possible Physical Realizations}

\vskip \baselineskip

It seems difficult from the experimental point of view to perform
measurements on a system of highly diluted nonionized hydrogen atoms, since
one has to avoid the atoms from reaching ionization (which is favored at
high temperatures), as well as from
achieving combination, ${\rm H} + {\rm H} \rightarrow {\rm H}_{2}$ 
(which becomes enhanced at low temperatures). In the following discussion
we propose two possible physical situations in which hydrogen atoms may be
found in the above-mentioned long-living quasi-equilibrium state. In both
cases, the duration of such a state is estimated in real time. 

\vskip \baselineskip
\noindent
\underline{A. Gas of Hydrogen Atoms}

Let us assume the viability, from the experimental point of view, for 
generating a gas of nonionized hydrogen atoms at low 
temperatures
(as compared with the corresponding ionization temperature). We shall also
assume that the following conditions are satisfied:

(i) The combinations, ${\rm H} + {\rm H} \rightarrow {\rm H}_{2}$, are
negligible. In fact, there are experimental techniques for such a purpose,
in which pairs of atoms in the so-called ``spin-polarized state'', can not
produce bound states 
\cite{silverawalraven,silverareview,walravenreview};   

(ii) The temperature $T$ is sufficiently low in such a way that most of the
atoms that compose the gas are in the ground state;

(iii) A given hydrogen atom can only change its state through collisions
with other atoms (we will assume that the linear dimension of the box
enclosing the gas is much larger than its mean free path, in such a way
that collisions with the walls may be neglected). 
An atom in its ground state may experience several
collisions before excitation (actually, such an atom should undergo a
collision with a sufficiently energetic atom in such a way to absorb an
energy greater than $3R/4$ for a transition to occur).   

Therefore, the mean time between
two successive collisions, $\tau$,
may be obtained from standard kinetic-theory calculations 
(see, e.g., ref. \cite{reif}, chapter 12), 

$$
\tau \approx  {1 \over 16 \sqrt{\pi} \  na^{2}}
\left( {m \over k_{B}T} \right)^{1/2}~, 
\eqno(13) 
$$

\vskip \baselineskip
\noindent
where $n$ represents the density of atoms in the gas, $m$ is the mass of
the hydrogen atom, and $a$ stands for the Bohr's radius. 
In the present physical system, a given atom can only change its energy
through collisions with other atoms; this situation is mimicked in the 
Monte Carlo simulation, where at each
step there is a finite probability for the occurence of a given energetic 
transition. 
In order to establish a connection between the duration of the
quasi-equilibrium states of Fig. 1 (given in Monte Carlo steps) with real
time, we shall propose the crude -- but very suggestive -- correspondence: 
$\tau \equiv 1$ Monte Carlo step. With this assumption, one gets that the
duration (in real time) of a hydrogen atom quasi-equilibrium state,
may be written as, 

$$
t_{\rm real} = t_{\rm MS} \ \tau \approx  
{1 \over 16 \sqrt{\pi} \  na^{2}}
\left( {m \over k_{B}T} \right)^{1/2} \exp [2 R/(k_{B}T)] ~, 
\eqno(14) 
$$

\vskip \baselineskip
\noindent
where we have considered the fitting parameter $b=2$ [cf. Eq. (5)]. 
It should be stressed that, at low temperatures, the exponential growth 
dominates completely, with the multiplicative factors in Eq. (14) becoming 
irrelevant. Therefore, the hypothesis above for the connection of Monte
Carlo steps with real time may be softened in such a way that any linear
relationship between a given number of successive collisions and another
number of Monte Carlo steps would lead to the same exponential-growth
behavior of Eq. (14).
 
One sees that
the above duration time depends on two parameters, namely, the density of
atoms and the temperature, i.e.,   
$t_{\rm real} \equiv t_{\rm real}(n,T)$. We have estimated 
$t_{\rm real}(n,T)$ for the typical values 
$n=10^{23} {\rm atoms}/{\rm m}^{3}$
and 
$\theta=0.10$ ($T \approx 15797$ K), 
$\theta=0.05$ ($T \approx 7898$ K), 
$\theta=0.03$ ($T \approx 4739$ K), and  
$\theta=0.01$ ($T \approx 1580$ K), 
which yielded, respectively, the duration times,
$t_{\rm real} = 5.35$ seconds, 
$3.67 \times 10^{9}$ seconds,
$1.81 \times 10^{21}$ seconds 
($\approx 5.73 \times 10^{13}$ years), and 
$2.52 \times 10^{79}$ seconds 
($\approx 8.0 \times 10^{71}$ years). 
One notices tremendous duration times for the lowest
temperatures. 

It is important to remind that the above results have not taken into
consideration the degeneracy of the energy spectrum, which should 
contribute to decrease the duration of the quasi-equilibrium states (since
the degeneracy increases the number of possible transitions taking place 
in the Monte Carlo process). 
However, the enormous estimates found above,
for low temperatures, should not be altered significantly due to the
degeneracy of the energy spectrum. 

\vskip \baselineskip
\noindent
\underline{B. Hydrogen Atoms in a Photon Bath}

It is well known that long-living nonionized hydrogen 
atoms exist in very low concentrations and at very low temperatures
(typically 3 K), in intergalactic media. These atoms are in direct
contact with photons, in such a way that transitions between states occur
through photon emission and absorption. 

In what follows, we will present a
crude estimate of a lower bound for the duration (in real time) of the
above-mentioned quasi-equilibrium state for a hydrogen atom in a 
medium such as
the intergalactic media. For that we will consider:
 
(i) A nonionized hydrogen atom in contact with a photon bath at a
temperature $T$; 

(ii) The temperature $T$ sufficiently low for the atom to be found
initially in its ground state; 

(iii) The velocity of the atom negligible with respect to the velocity
of light.

Let us define $t_{1 \rightarrow n}(\theta)$ as the average time that 
the atom takes
to absorb a sufficiently energetic photon [with an angular
frequency $\omega \ge \Lambda=(3R)/(4\hbar)$], in such a way as 
to perform a
transition from the ground state to an excited state characterized by a
quantum number $n$ ($n>1$), for a given value of $\theta$. Obviously, 
$t_{1 \rightarrow n}(\theta)$ represents a lower-bound estimate for the
duration of the quasi-equilibrium state, since the atom may return afterwards 
to its ground state (this may occur with a large
probability, since we  are assuming low temperatures). Let us then 
consider the average number of photons per unit volume (including both
directions of polarization), with angular frequency $\omega \ge \Lambda$,  

\vspace{-5mm}

$$
n_{\omega \ge \Lambda}(\theta) = {8 \pi \over (2 \pi c)^{3}}
\int_{\Lambda}^{\infty} {\omega^{2} d\omega \over 
\exp [(\hbar \omega) / (R \theta)]-1}~,
\eqno(15) 
$$

\vskip \baselineskip
\noindent
where we have used the variable $\theta=k_{B}T/R$. For the temperature
range of interest, one has that $\exp [(\hbar \omega) / (R \theta)] \gg 1$,
in such a way that the above integral may be calculated easily,

\vspace{-5mm}

$$
n_{\omega \ge \Lambda}(\theta) \cong {8 \pi \over (2 \pi c)^{3}}
\exp \left( -3 \over 4\theta \right)
\left[ 2 \left( {R\theta \over \hbar} \right)^3
+ 2\Lambda \left( {R\theta \over \hbar} \right)^2
+\Lambda^{2} {R\theta \over \hbar} \right]. 
\eqno(16) 
$$

\vskip \baselineskip
\noindent
Using the result above one may estimate the number of photons with energy 
$\hbar \omega \ge 3R/4$ in the volume of the hydrogen atom (to be 
considered herein as $4 \pi a^{3}/3$, where $a$ represents Bohr's radius). 
The maximum time that these photons spend within the
volume of the atom is given by $2a/c$. Therefore, the average number of such
photons in the volume of the hydrogen atom, per unit time, may be written
as 

\vspace{-5mm}

$$
N_{\omega \ge \Lambda}(\theta) = n_{\omega \ge \Lambda}(\theta) 
\left( {c \over 2a} \right) \left( {4 \over 3}\pi a^{3} \right)
\cong {2 a^{2} \over 3 \pi c^{2}}
\exp \left( -3 \over 4\theta \right)
\left[ 2 \left( {R\theta \over \hbar} \right)^3
+ 2\Lambda \left( {R\theta \over \hbar} \right)^2
+\Lambda^{2} {R\theta \over \hbar} \right]
\eqno(17) 
$$

\vskip \baselineskip
\noindent
From this result one calculates the
average time for the hydrogen atom to absorb a photon with sufficient 
energy to perform the transition $1 \rightarrow n$,
$t_{1 \rightarrow n}(\theta)N_{\omega \ge \Lambda}(\theta) = 1$, i.e., 
$t_{1 \rightarrow n}(\theta)=[N_{\omega \ge \Lambda}(\theta)]^{-1}$. It is
important to notice that, similarly to what happens for the duration of the
quasi-equilibrium state -- measured previously in Monte Carlo steps
[cf. Eq. (5)] -- the lower bound $t_{1 \rightarrow n}(\theta)$ (in real time
units) also follows an Arrenhius law,

\vspace{-5mm}

$$
t_{1 \rightarrow n}(\theta) \sim \exp [3/(4\theta)] = \exp [3 R/(4k_{B}T)].
\eqno(18) 
$$

\vskip \baselineskip
\noindent
However, as expected, the factor multiplying $(1/\theta)$ in Eq. (18) 
is smaller than the one found in Eq. (5). 
One should notice that there may be alternative ways to obtain 
$N_{\omega \ge \Lambda}(\theta)$ through the knowledge of 
$n_{\omega \ge \Lambda}(\theta)$, as done in Eq. (17), e.g., by introducing
a different time dependence in Eq. (17). However, the most important
behavior, i.e., the Arrenhius law of Eq. (18), remains unchanged by using a
different calculation for $N_{\omega \ge \Lambda}(\theta)$.
It is important to mention that, in the present example there was no need
for a connection between real time and Monte Carlo steps, as
done in the previous case. Even though, the low-temperature 
exponential-growth behavior appeared again; 
this result supports the hypothesis carried in the previous
case for such a connection. 

Lets us now consider two typical
examples for the lower bound $t_{1 \rightarrow n}(\theta)$, namely
$\theta=0.01$ ($T \approx 1580$ K) and 
$\theta=0.0001$ ($T \approx 15.80$ K), which lead,
respectively, to the colossal times (even when compared with the age of 
the universe of about $1.5 \times 10^{10}$ years),   
$t_{1 \rightarrow n}(0.01) \cong 3.52 \times 10^{16}$ years and 
$t_{1 \rightarrow n}(0.0001) \cong 1.57 \times 10^{3243}$ years. Since the
duration of the quasi-equilibrium state (in real time units) should be 
much larger than 
$t_{1 \rightarrow n}(\theta)$, it becomes evident the treatment of the
quasi-equilibrium state considered herein, as an effective equilibrium 
state for low temperatures.

\vskip 2 \baselineskip
\noindent
{\large\bf 6. \quad Conclusion}

\vskip \baselineskip

We have shown that the hydrogen atom may live in 
a quasi-equilibrium
state, at low temperatures -- in comparison with its corresponding 
ionization temperature -- whose duration increases exponentially 
as the temperature decreases. By considering such a quasi-equilibrium state as
an effective equilibrium state, we have calculated, for the first time (to
our knowledge), thermodynamic properties within the BG
statistical mechanics. For that, we have proposed a modified formalism
(whose results are very close to those obtained through numerical 
simulations using
the BG weight factor in the quasi-equilibrium state, at low
temperatures), characterized by a regularized partition function.  
It should be stressed that such an approximation is
supposed to be valid up to temperatures corresponding to about 
$6 \% $ of the ionization energy. Since the ionization energy $E^{*}$ of 
the hydrogen atom is extremely high ($E^{*} \approx 13.61$ eV, i.e.,
$E^{*}/k_{B} = 1.579 \times 10^{5}$ K), our approximation
should work well up to temperatures $10^{4}$ K.
It is important to mention the broad interest of the above analysis, which
applies to many atoms,
molecules, composite particles, and other similar systems, characterized
by: (i) upper bounds, preceded by a quasi-continuum of energy levels, in
their energy spectra; (ii) large gaps separating the ground and
first-excited states.
Obviously, experimental investigations are highly desirable in order to
test the validity of these results. 

%\vskip 2\baselineskip
\newpage

{\large\bf Acknowledgments}

\vskip \baselineskip
\noindent
We thank Constantino Tsallis for fruitfull
discussions. The partial financial supports from CNPq and Pronex/MCT
(Brazilian agencies) are acknowledged. F.D.N. would like to thank 
Centro Brasileiro de
Pesquisas F\'{\i}sicas (CBPF), where this work was developed, for
the warm hospitality. 

%\newpage
%\vskip 2\baselineskip
\vskip \baselineskip

\end{document}